\documentstyle[12pt,epsf]{article}
\newcommand{\m}{\medbreak}
\newcommand{\no}{\noindent}
\newcommand{\EQ}{\begin{equation}}
\newcommand{\eq}{\end{equation}}
\newcommand{\EQA}{\begin{eqnarray}}
\newcommand{\eqa}{\end{eqnarray}}

\newcommand{\z}{$Z^{'}\;$}
\newcommand{\ZP}{$Z^{'}\;$}

\newcommand{\ppv}{\mbox{$\vec{p}\vec{p}\; $}}
\newcommand{\nnv}{\mbox{$\vec{n}\vec{n}\; $}}
\newcommand{\pp}{\mbox{${\vec{p}}p\; $}}
\newcommand{\nn}{\mbox{${\vec{n}}n\; $}}
\newcommand{\alpp}{\mbox{$A_L(\vec{p}p)\ $}}

\newcommand{\alnn}{\mbox{$A_L(\vec{n}n)\ $}}

\begin{document}
\begin{titlepage}
\vspace{0.2in}
\vspace*{1.5cm}
\begin{center}
{\large \bf Constraints on leptophobic gauge bosons with polarized
 neutrons
 and protons at RHIC} \\
\vspace*{0.8cm}
{\bf P. Taxil}$^a$, {\bf E. Tu{\u g}cu}$^b$ and {\bf J.-M. Virey}$^a$  \\ \vspace*{1cm}
$^a$Centre de Physique Th\'eorique$^{\ast}$, C.N.R.S. - Luminy,
Case 907\\
F-13288 Marseille Cedex 9, France\\ \vspace*{0.2cm}
and \\ \vspace*{0.2cm}
Universit\'e de Provence, Marseille, France\\ \vspace*{0.5cm}
$^b$Galatasaray University, \c 
C\i ra\u gan Cad. 102, Ortak\"oy 80840-\.Istanbul, 
Turkey\\
\vspace*{1.8cm}
{\bf Abstract} \\
\end{center}
We explore the sensitivity and the physics interest
of the measurement of parity-violating spin asymmetries in one-jet
production in the presence of a new leptophobic
neutral gauge boson \ZP, within polarized hadronic
collisions at the BNL RHIC. We focus on polarized neutron collisions which could
be achieved in 
a realistic upgrade of the RHIC-Spin program.
We show that, in case of a discovery, 
a compilation of the information coming
from both polarized \ppv and \nnv collisions should constrain the number
of Higgs doublets and the presence or absence of trilinear fermion
mass terms in the underlying model of New Physics.
 \\

\vfill
\begin{flushleft}
PACS Numbers : 12.60.Cn; 13.87.-a; 13.88.+e; 14.70.Pw\\
Key-Words : New Gauge bosons, Jets, Polarization, Neutron Collisions.
\m\no
Number of figures : 4\\

\m\no
November 2001\\
CPT-2001/P.4263\\
\m\no
anonymous ftp or gopher : cpt.univ-mrs.fr

------------------------------------\\
$^{\ast}$Unit\'e Propre de Recherche 7061
 \\
E-mail : taxil@cpt.univ-mrs.fr ; tugcu@gsunv.gsu.edu.tr ; virey@cpt.univ-mrs.fr
\end{flushleft}
\end{titlepage}


\section{Introduction}

The addition of an extra $U(1)'$ gauge factor to the $SU(3)\times SU(2)\times U(1)$
structure is one of the simplest extensions of the Standard Model (SM).
When the symmetry breaking of this extra factor occurs at a scale close to
the electroweak scale, one obtains a new neutral gauge
boson $Z'$ in the particle spectrum, 
at a mass accessible to forthcoming experiments.\\

The strongest experimental constraints on such $Z'$ models come from experiments
which analyze some processes involving leptons, either in the initial state
or/and in the final state. For instance, the constraints coming from LEP, HERA or
the Drell-Yan process at Tevatron are complementary and provide some bounds on the
$Z'$ mass of the order of $600-700\; GeV$ for canonical models \cite{PDG}, 
the precise values
depending on the specific model and the relevant process involved in the analysis.
At the same time the $Z - Z^{'}$ mixing angle $\theta_{Z-Z^{'}}$ is constrained to 
be very small.
However, when the $Z'$ has zero or very small direct couplings to leptons 
(leptophobia), the above processes are irrelevant and one has to turn to pure hadronic
channels to provide some constraints \cite{PDG}.

The existence of relatively light leptophobic gauge bosons
is an attractive possibility, both for phenomenology and from theoretical
arguments. Recent papers advocated a weak-scale supersymmetry (SUSY)
scenario in non-minimal SUSY models with an additional extra $U(1)^{'}$.
The corresponding $Z'$ could be "light" : $M_{Z'}<1.5\, TeV$. More precisely,
a class of models driven by a large
trilinear soft SUSY breaking term, prefer the range $M_Z <M_{Z'}<400\, GeV$, along
with a very small mixing with the standard $Z$.
This particular scenario is only allowed
if the model exhibits leptophobic couplings \cite{Cvetic}.
On the other hand, other models display or can accommodate leptophobia : 
some are string inspired \cite{Lykken,Babu,LopezNanopoulos},
others are non-SUSY \cite{NoSusy,GG}
(for a more complete set of references one can consult our paper \cite{PTJMVZ'}). 
Furthermore, 
in many models an asymmetry in the left and right-handed couplings of the \ZP to
light quarks is preferred or at least allowed.\\

In a previous paper \cite{PTJMVZ'}, we have shown that the measurement of 
Parity Violating (PV) spin effects in the production of jets
from  hard collisions of polarized hadrons, could be a way to get a
handle on this elusive  leptophobic $Z'$ boson. \\
The situation of interest is the one at Brookhaven National Laboratory
where the RHIC machine is operating mainly as a heavy-ion collider  
but will be used part
of the time as a polarized proton-proton (\ppv) collider.
The RHIC Spin Collaboration (RSC) has performed a first run
during the year 2001 , with
polarized protons, an energy $\sqrt s = 200 $ GeV and
a luminosity of a few $10^{30} cm^{-2}s^{-1}$.
Around 2003, it is expected to reach $\sqrt s = 500 $ GeV and 
${\cal L}\ =\, 2. 10^{32} cm^{-2}s^{-1}$ \cite{BNL01}
allowing an exposure
of 800 $pb^{-1}$ in only four months of running.
The physics program of the collaboration has been reviewed recently
in ref.\cite{BSSW} where many references can also be found 
(see also \cite{BNL01}).
This program will allow first some precise measurements
of the polarization of the gluons, quarks and sea-antiquarks
in a polarized proton. This will be done thanks to well-known
Standard Model processes : direct photon, $W$ and $Z$ production, 
Drell-Yan pair production, heavy-flavor production
and the production of jets.
The helicity structure of perturbative QCD will be thoroughly
tested at the same time with the help of Parity Conserving
(PC) double spin asymmetries.

Concerning new physics, it has been noticed that
non-zero $CP$-violating asymmetries can be generated from various mechanisms
going beyond the SM \cite{Kovalenko,Grad,Atwood}. 
On the other hand, 
the production of high $E_T$ jets from polarized protons
could allow to pin down
a possible new weak interaction between quarks, provided that parity is 
violated in the subprocess \cite{MTPenn,PTJMVCI,PTJMVW',PTJMVZ',PTJMVup}.
In the case of a simple phenomenological PV Contact Term, a search
strategy based on the polarized RHIC can be competitive with
conventional searches at the Tevatron, or even better 
\cite{PTJMVPRD,PTJMVup}.\\

Indeed, the production of jets is largely dominated by QCD, which is
a parity conserving theory. However a Standard PV spin asymmetry 
in jet production should be present from tiny QCD-electroweak 
interference effects, namely the interference between the one-weak boson
exchange amplitude and the one-gluon exchange amplitude since, at high 
$E_T$, the process is dominated by $q-q$ scattering.
The magnitude and sign of this Standard PV asymmetry can be safely
estimated from well-known subprocess amplitudes and from our
knowledge of the polarized quark distributions in a polarized protons.
Note that polarized gluons distributions (which are poorly known) are
irrelevant in this process at least at Leading Order (LO).
Therefore, a net deviation from the small expected Standard Model
asymmetry could be a clear signature of the presence of a new force
belonging to the quark sector with a peculiar chiral structure.

Models with leptophobic $Z'$s are obviously good candidates to consider in this
context. The study presented
in \cite{PTJMVZ'} has shown that, in order to detect a non standard effect
in \ppv collisions at RHIC,
besides the necessity of leptophobia plus a low mass, the $Z'$ boson must exhibit
an asymmetry in the left and right couplings to $u$ quarks since $u$ quarks
dominate in \ppv collisions. 
Fortunately, the existence of such PV couplings for
$u$ quarks is a prediction of several leptophobic models constructed up
to now \cite{Babu,LopezNanopoulos,NoSusy,GG}.

Conversely, the PV nature of the $Z'$ couplings to $d$ quarks is much more
model dependent. Indeed, it depends on the symmetry breaking scenarios
and on the scalar potential assumed for the models. More precisely, we will see that
the PV properties of the $d$ couplings are directly connected to the number
of Higgs doublets involved in the model and to the presence or absence of tri-linear
fermion mass terms in the Yukawa lagrangian. So, in case of a
discovery, the measurements of these $d$ couplings,
or at least the test of their PV nature, should provide a unique information
on the scalar sector of the underlying theoretical model of New Physics.\\

Unfortunately, within \ppv collisions at RHIC the $Z'$ amplitudes involving  
 $d$ quarks in the initial and final states are completely hidden by the $u$ quark
contributions. However, a particular feature of the RHIC as a heavy ion collider, is
to be able to accelerate polarized $^3He$ nuclei, which could mimic high energy
polarized neutrons. Indeed, the Pauli exclusion principle implies that the polarized
$^3He$ nuclei carries essentially the spin of the neutron since the spins of the
protons are in opposite directions.

This possibility has been considered by the RSC and it is expected to get some
polarized beams of ``neutrons'' of relatively good quality \cite{Courant}.
Therefore, in the following, we consider polarized \nnv
collisions at RHIC in order to explore which kind of information could be
obtained on the $d$-couplings and on the scalar sector of the new theory.\\

In section two, we present the models and the different scalar structures
which are considered in our analysis.
In section three, we present the definition of the spin 
dependent observable that we consider, 
we summarize our calculations and we give the limits on the parameter
space which could be achieved at RHIC in the case of the various models,
within \nnv collisions. 
In the last section, we show a combined analysis of the information which could be
provided by both \ppv and \nnv collisions, in case of a discovery, 
and therefore the constraints
that might be obtained on the Higgs sector of the theory .\\

\section{Classification of the models}

\vspace{1mm} 
\noindent

The interactions between a new neutral vector gauge boson \z and up and down-type 
quarks are described by the following lagrangian: 

\begin{eqnarray}
{\cal L}_{Z^{'}}&=&\kappa\frac{g}{2cos\theta_W}\sum_q 
Z^{'\mu}\bar{q}\gamma_\mu\left[{C^q_L}\left(1-\gamma_5\right)+{C^q_R}
\left(1+\gamma_5\right)\right]q
\end{eqnarray}
where $C^q_{L,R}$ are the couplings to left and right handed quarks for each 
given quark flavor $q$ and the parameter $\kappa ={g_{Z^{'}}}/{g_Z}$
being of order one.
We restrict our discussion
to the light quark flavors $u$ and $d$ since only a $Z'$ which couples
to these light quarks may give some observable effects at RHIC energies.\\
In what follows, we will 
concentrate on leptophobic \z models of relatively light masses with chiral 
couplings to quarks. 
We refer the reader to ref.\cite{PTJMVZ'} and to the original literature for
more details on the theoretical motivations and on the
underlying structures of each model.\\

At first, we consider an approach similar to the one of Georgi and Glashow 
\cite{GG} to determine the general conditions imposed by gauge invariance, 
leptophobia and symmetry breaking on the $U(1)'$ charges. Then,  each different 
general situation will be illustrated by a specific model. \\
First of all, gauge invariance under the SM group $SU(2)_L$ imposes the
universality of the left-handed couplings :
\EQ
{C_L^u}\;=\;{C_L^d}\; \equiv\; C_L
\eq
\no Therefore, in the following, we will suppress the flavour indice on the left-handed
couplings.\\

Initially we can assume that all SM fermions 
acquire their masses via the trilinear mass terms present in the Yukawa
lagrangian :

\begin{equation}
{\cal L}_Y\;=\;h_u\bar{Q}H_uu_R\;+\;h_d\bar{Q}H_dd_R\;+\;h_l\bar{L}H_le_R
\end{equation}
\no where $Q$ is a quark doublet, $L$ a lepton doublet, $u_R$, $d_R$
and $e_R$ are right-handed singlets.
 $H_{u,d,l}$ represent the
 corresponding Higgs doublets and $h_{u,d,l}$ are Yukawa coupling matrices.
For supersymmetric models, the structure is formally the same on condition 
that one replaces
the potential by the superpotential and the fields by the superfields.\\

Gauge invariance under the new $U(1)^{'}$ gauge group associated with the $Z'$, 
imposes that the sum of the $U(1)^{'}$ charges $Q^{'}$ for each term is zero :

\EQ
Q^{'}(H_u)\;-\;Q^{'}(Q)\;+\;Q^{'}(u_R)\;\; =\;\; 0 
\eq
\EQ
Q^{'}(H_d)\;-\;Q^{'}(Q)\;+\;Q^{'}(d_R)\;\; =\;\; 0 
\eq
\EQ
Q^{'}(H_l)\;-\;Q^{'}(L)\;+\;Q^{'}(e_R)\;\; =\;\; 0 
\eq
\no The $U(1)'$ charges of the fermions are directly related to their chiral couplings :
$Q^{'}(Q)=C_L$, $Q^{'}(L)=C_L^{e,\nu}$ and $Q^{'}(f_R)=C_R^f$.\\

From eq.(6) the condition of leptohobia $Q^{'}(e_L)=Q^{'}(e_R)=0$, 
forces the charge of the Higgs doublet coupling to the lepton field to be zero :

\begin{equation}
Q^{'}(H_l)\;=\;0\;
\end{equation}

Given these assumptions, we will describe now three different scalar
structures implying different properties for the right-handed couplings
of $d$ quarks to the \ZP. Of course, one may be surprised by this particular approach
where the $Q'$ charges seem to be put by hand instead of taking a specific
model where these charges are fixed and where anomaly cancellation
is fulfilled thanks to the presence of exotics. Here we are not interested
in these exotics since we can assume safely that their masses are sufficiently 
high to avoid detection at existing colliders, including RHIC. 
Moreover,
we want to choose an approach which is as most as possible model independent. 
Since, at RHIC, we
can test  the PV structure of $d$ quark couplings, we just quote which choice
of scalar structure, independently of the choice of a particular new gauge theory,
implies a modification in this PV structure.\\

\no {\bf  $\bullet$ Structure I : 2HDM}\\

A first interesting case appears when the Higgs
doublet $H_d$ which generates the masses of $d$-type quarks is identical to the one 
which yields (charged) lepton masses. This structure corresponds to
the Two Higgs Doublets Models (2HDM) and it can be achieved for special values
of the $Q'$ charges \cite{GG}. 
In this case, $H_l\equiv H_d$ and we have 
from eq.(7) :
\EQ
Q^{'}(H_d)=0
\eq 
\no This implies from eq.(5) $Q^{'}(Q)= Q^{'}(d_R)$ or, in terms of the couplings :
\begin{equation}
{C_R^d}\;=\;{C_L}
\end{equation}

\no We see that the $d$ quark couplings are {\it vector-like}
which means that Parity is conserved in this quark sector. 
This is the main characteristic of the models displaying
this structure that we call {\it structure I} from now.\\

Conversely, if we want the remaining Higgs doublet $H_u$ to play a role in the
symmetry breaking of the $U(1)'$ symmetry, then it must be charged under
$U(1)'$ in order to acquire a vacuum expectation value that breaks the 
$U(1)'$ symmetry.
So, we take $Q'(H_u) \neq 0$ which implies for the couplings (see eq.(4)) :
\EQ \label{10}
{C_R^u}\;\neq \; {C_L}
\eq

From this equation, we see that the $u$ quark couplings cannot be vector-like.
Hence, Parity will be violated, except for the peculiar axial case
where ${C_R^u}\; = \; -{C_L}$.

This remark and  eq.(\ref{10}) are also valid for the remaining structures that we will
consider, {\it i.e.} we always assume that $H_u$ is playing a role in the breaking of
the $U(1)'$ symmetry. Note that additional scalars, singlets of $SU(2)_L$, can also be
present in the breaking scenario but they have no impact on the PV
nature of the $d$ couplings. \\

An explicit model which can be an illustration of this {\it structure I} is the 
supersymmetric "$\eta$-kinetic model" of Babu, Kolda and March-Russell 
\cite{Babu}, whose properties are Parity Conservation  for $d$ quarks
and PV for $u$ quarks.
We call it {\it Model B} to remain coherent with reference 
\cite{PTJMVZ'}:

\begin{equation}
Model\;B\;:\;\;\;\;\;\;\;\;C^u_L\;=\;C^d_L\;=\;C^d_R\;=\;
-\frac{1}{2}C^u_R\;=\;-\frac{5}{18}sin\theta_W
\end{equation}

\no Note that, at a variance with the notations of 
ref.\cite{Babu} the usual factor of Grand
Unification Theories  ($\sqrt{\frac{5}{3}}sin\theta_W$) has been included in the
$C^q_{L,R}$  couplings to maintain $\kappa$ of order one.\\

\no {\bf $\bullet$ Structure II :}\\

The first extension of the previous scalar structure is achieved when we allow the 
third Higgs doublet $H_l$ to be different from $H_d$. In this 
case, leptophobia does not provide anymore a direct relation among the 
couplings $C_L$ and $C^d_R$, since eq.(5) and eq.(6) are now completely disconnected.
For {\it structure II} we still consider that eq.(5) is 
valid, {\it i.e.} that $d$ quarks acquire their masses from a trilinear mass term.
Conversely, we don't make any assumption on the form of the (charged) lepton 
mass term which plays no role in the following discussion.\\

Nevertheless, in many extensions of the 
SM, particularly in most of the leptophobic 
\z models, it is assumed that the symmetry breaking is driven by the two vacuum 
expectation values of the $H_u$ and $H_d$ Higgs doublets which are 
of the same order ($v_{H_u}\simeq v_{H_d}$) \cite{GG}. Indeed, the 
constraints 
from the electroweak precision data impose that the $Z-$\z mixing angle, 
$\theta_{Z-Z^{'}}$ should be very small. This requires that \cite{GG}:

\begin{equation}
g_{Z^{'}}^2\left|\;v_{H_u}^2Q^{'}\left(H_u\right)\;-\;v_{H_d}^2Q^{'}\left(H_d
\right)\;\right|\;<<\;g_Z^2v^2
\end{equation}
With $v_{H_u}\simeq v_{H_d}$, this expression leads us to assume that 
$Q^{'}(H_u)$ and $Q^{'}(H_d)$ are also of the same order and have the same 
sign : 
\EQ
Q^{'}(H_u)\simeq Q^{'}(H_d)
\eq

\no Using the condition of $U(1){'}$gauge invariance 
of the trilinear $u$ and $d$ mass terms (eq.(4) and eq.(5)), 
we get the following relation among the chiral couplings :

\begin{equation}
{C_L}\;-\;{C^u_R}\;\simeq\;{C_L}\;-\;{C^d_R}
\end{equation}
Thanks to the $SU(2)_L$ gauge invariance this gives :
\begin{equation}
{C^u_R}\simeq {C^d_R}\;
\end{equation}

It clearly means that the left or right handed dominance is the same for 
$u$ quarks and for $d$ quarks. An example of such models, 
is given by the
following right-handed model we call {\it Model C }\footnote{This model is analogous
to the second model of \cite{GG}, but we have changed the precise values
of the $U(1)'$ charges in order to have $\kappa \simeq 1$.} :

\begin{equation}
Model\; 
C\;:\;\;\;\;\;\;\;\;{C_L}\;=\;0,\;\;\;\;\;{C^u_R}\;=\;{C^d_R}
\;=\;\frac{1}
{3}
\end{equation}

Note that some authors consider eq.(14) and eq.(15) simply
as orders of magnitude.
For example, the model presented in \cite{NoSusy} (which is also the first
model of \cite{GG}) fits into this scalar {\it stucture II }, but the
$U(1)'$ charges are : $Q'(H_u)=-3$, $Q'(H_d)=-2$,  $Q'(Q)=C_L=-1$,
$Q'(u_R)=C_R^u=2$, and $Q'(d_R)=C_R^d=1$. Therefore, we get PV couplings for
$u$ quarks but axial couplings for $d$ quarks.\\
Then, in order to be conservative, one can say that a characteristic of the models
with the scalar {\it stucture II } is that they cannot yield
a left handed (right handed) dominance for $u$ quarks and a right handed 
(left handed) dominance for  $d$ quarks at the same time.\\

\no {\bf $\bullet$ Structure III :}\\

Finally, we can consider the non-minimal scalar structure provided by string derived 
models as the ones considered by Cveti{\v c}, Langacker and collaborators
 \cite{Cvetic}.
A peculiarity of these string derived models is 
that trilinear mass terms appear naturally for 
$u$ or $d$-type quarks but not for both \cite{Faraggi}.
A correct prediction for the top quark mass is done \cite{Faraggi2}
if one takes a trilinear mass term for the top quark \cite{Faraggi}.
This choice is made in \cite{Cvetic}. In these scenarios, $d$-type quarks
and charged leptons acquire their mass thanks to nonrenormalizable terms
({\it i.e} not trilinear).

If there is no trilinear mass term in the theory for $d$-quarks, 
eq.(5) is no longer valid. Therefore, the 
chiral couplings of the $d$ quarks are completely free. It means that
we can have PC couplings (vector-like as in {\it structure I}, or axial), 
or PV couplings
with a left-handed or a right-handed dominance, the same as for $u$
quarks (this is similar to {\it structure II} ), or on the contrary in opposition 
to the case of $u$ quarks. This situation will characterize {\it structure III}.

We have chosen the following phenomenological {\it Model 
D} to illustrate this last possibility :

\begin{equation}
Model\;D\;:\;\;\;\;\;\;\;\;{C_L}\;=\;\frac{1}{3},\;\;\;\;\;{C^u_R}\;=
\;0,\;\;\;
\;\;{C^d_R}\;=\;\frac{2}{3}
\end{equation}

In addition, the flipped $SU(5)$ model of Lopez and Nanopoulos \cite{LopezNanopoulos} 
is another good example of  {\it structure III}, but now with axial 
couplings for $d$ quarks. This model, that we call {\it Model A} from reference 
\cite{PTJMVZ'}, is characterized by :

\begin{equation}
Model\;A\;:\;\;\;\;\;\;\;\;{C_L}\;=\;-\;{C^d_R}\;=\;\frac{1}{2\sqrt{3
}},\;\;\;
\;\;{C^u_R}\;=\;0
\end{equation}
These couplings imply that Parity is maximally violated in the $u$-quark 
sector whereas it is conserved in the $d$ quark sector because of the purely 
axial character of the couplings.\\

\section{Observables and Results}

We concentrate on the 
inclusive single jet production process $\vec{n}n\rightarrow jet+X$, where
the polarization of only one neutron is necessary to define
the single helicity PV asymmetry :
\EQ
A_L \; =\; {d\sigma_{(-)}-d\sigma_{(+)}\over 
d\sigma_{(-)}+d\sigma_{(+)}}
\eq
\no
where the signs $\pm$ refer to the helicity of the polarized neutron. 
The cross section $d\sigma_{(\lambda )}$ means the one-jet
production cross section estimated at  
some $\sqrt{s}$
for a given jet transverse energy $E_T$, integrated over a pseudorapidity interval 
$\Delta \eta \,$ centered at $\eta\,=\,0$.\\

In fact, both $^3He$ beams could be polarized in principle, giving access to doubly
polarized neutron collisions $\vec{n}\vec{n}\rightarrow jet+X$
and to double-helicity asymmetries (for a review on definitions
and calculations of spin observables one can consult ref.\cite{BRST}). Then the
statistical significance is increased but a similar amount of information 
is obtained on the chiral and  scalar structures. We prefer to be conservative
considering that only one beam will be polarized.\\

Concerning the value of $\sqrt s$,
at RHIC, the charged nucleons (protons) of a nuclei 
are accelerated up to 
energies of $E_p=250-300\;GeV$ per nucleon. At first, the machine will run with 
$E_p=250\;GeV$, it is the reason why we have taken a center mass energy
$\sqrt{s}=500\;GeV$ for  $\vec{p}\vec{p}$ collisions \cite{PTJMVZ'}. 
A $^3He$ nuclei, being accelerated, will get the 
total energy of its two protons. The neutron will be able to reach only one 
third of this energy, which means that the center of mass energy will be reduced
for $n - n$ collisions. 
To be sensitive to a possible New Physics effect, it is necessary to  
run at the highest possible energy ({\it i.e.} $E_p=300\;GeV$). Hence, we have 
$E_{^3He}=600\;GeV$ and $E_n=200\;GeV$, \nnv (or ${\vec n}n$) collisions reaching 
an energy  $\sqrt{s}=400\;GeV$ in the center of mass, value that we take in the
following. 
Concerning the integrated luminosities,  we have taken the same values as in
\cite{PTJMVZ'}, namely $L_1=800\, pb^{-1}$ and $L_2=3.2\, fb^{-1}$ for practical
comparison with the \ppv results, even if these values are certainly a little bit
optimistic. However, if we take seriously the possibility of some luminosity
upgrades at RHIC in the future \cite{Saito}, these numbers become perfectly
realistic.\\

The dominant subprocess in the $E_T$ range that we consider, is 
quark-quark scattering. 
Concerning the Standard contribution,  $A_L$ is given by the
expression (in short notations): 
\EQ
\label{ALjet}
A_{L} \simeq  {1\over d\sigma}\sum_{i,j} \sum_{\alpha,\beta}
\int
\left(T_{\alpha,\beta}^{--}(i,j) - T_{\alpha,\beta}^{++}(i,j)
\right) 
\biggl[q_i(x_1)\Delta q_j(x_2) + \Delta q_i(x_1)q_j(x_) 
+ (i\leftrightarrow j) \biggl]
\eq \no
The
$T_{\alpha,\beta}^{\lambda_1,\lambda_2}(i,j)$'s are the matrix element squared
with  $\alpha$ boson and $\beta$ boson exchanges in a given helicity configuration
for the involved partons $i$ and $j$.  The expressions for the relevant 
$T_{\alpha,\beta}$'s at leading order (LO) 
are well-known, they can be found e.g. in ref.\cite{BGS}.
$\Delta q_i\ =\ q_{i+} - q_{i-}$ where 
$\, q_{i\pm} \equiv \; q_{i\pm}(x,\mu^2)$ 
are the distributions of 
the polarized quark of flavor $i$, either
with helicity parallel (+) or antiparallel ($-$) to the parent proton
helicity. Summing the two states one recovers  $q_{i+} + q_{i-} = q_i(x,\mu^2)$.
All these distributions are evaluated at the scale $\mu\, =\, E_T$. 
The
unpolarized cross section $d\sigma$ is dominated by QCD and must also include all the
relevant Electroweak+\ZP terms and their interference with QCD terms when it is
allowed by color rules.  Of course, the non-dominant $q(\bar q)g$ and $gg$ scattering
subprocesses have to be included in the part of the cross section which is
purely QCD.  The resulting standard $A_L$ is positive and increases with
the jet transverse energy $E_T$ as soon as $E_T$ is larger than the range
$E_T\approx M_{W,Z}/2$ (see the figures below). This is due to the increasing
importance of quark-quark scattering with respect to other subprocesses
involving gluons.
\m


If present, the leptophobic \z contributes to the quark-quark scattering process via
new  amplitudes which interfere with the single gluon exchange amplitude. 
It is straightforward to get these amplitudes from the
Standard ones involving the standard $Z$. One has also to add the very 
tiny Electroweak-\ZP interference which will not yield an observable effect.
As it was already pointed out in reference \cite{PTJMVZ'}, 95\% of the effect 
due to the new boson comes from $Z'$-gluon interference terms involving the 
scattering of identical quarks in the $t,u$-channels. In the case of ${\vec n}n$
 collisions
it corresponds essentially to the scattering of $d$ quarks.
This dominant contribution can be written as 
follows :

\begin{equation} \label{ALs}
A_{L}\cdot 
d\sigma\;\simeq\;F\int\left[C_L^2-{C^d_R}^2\right]\left[d\left(x_1,\mu^2\right)\Delta 
d\left(x_2,\mu^2\right)+
\Delta d\left(x_1,\mu^2\right)d\left(x_2,\mu^2\right)\right]_{neutron}
\end{equation}
where $F$ is a positive factor given by

\begin{equation}
F\;=\;\frac{32}{9}\alpha_s\alpha_z\hat{s}^2Re\left(\frac{1}{\hat{t}D^{
\hat{u}}_
{Z^{'}}}+\frac{1}{\hat{u}D^{\hat{t}}_{Z^{'}}}\right)
\end{equation}
where $\alpha_z=\alpha/sin^2\theta_Wcos^2\theta_W$ and 
$D^{\hat{t}(\hat{u})}_{Z^{'}}=\left(\hat{t}(\hat{u})-M^2_{Z^{'}}\right
)+iM_{Z^{'
}}\Gamma_{Z^{'}}$.\\

Note that the partonic part of eq.(\ref{ALs})
corresponds to the polarized and unpolarized $d$
quark distributions in a {\it neutron}. So, if we want to use an expression
with the more familiar definitions of quark distributions in a {\it proton}, 
from isospin symmetry,
we have to replace $d_{neutron}(x,\mu^2)$ and $\Delta d_{neutron}(x,\mu^2)$ by 
the functions $u(x,\mu^2)$ and $\Delta u(x,\mu^2)$, defined for a proton.
It means that the partonic part of eq.(\ref{ALs}) is positive since 
$\Delta u_{proton}$ is positive as  well-known. \\

We can remark that eq.(\ref{ALs}) allows us to predict easily the behaviour 
of the spin  asymmetry $A_L$ in the presence of a new \z contribution. 
Given the positivity of the factor $F$ and of the partonic part, the 
direction of a possible deviation from the SM  $A_L$ asymmetry
will be determined directly by 
the chiral couplings $C^d_{L,R}$, more precisely by the difference
$C_L^2-{C^d_R}^2$. Consequently, a model whose $d$ chiral 
couplings present a left (right) dominance will provide a positive (negative) 
deviation to the SM $A_L$ asymmetry.\\

In our LO calculations, all the contributions, dominant or not are included.
Concerning the partonic part we have used
the GRSV polarized parton distribution functions (pdf) \cite{GRSV}
along with the associated unpolarized pdf's. 
Remember that the uncertainties
due to the imperfect knowledge of the polarized pdf's will be reduced soon thanks to
the first part of the RHIC-Spin program itself \cite{BSSW}.

On the theoretical side some systematic uncertainties are coming from
the existence of higher order corrections to the SM prediction for $A_L$
and to the \ZP contribution itself.
Indeed, at NLO, several new contributions appear \cite{JMVBNL00}. 
However, the current prejudice is that spin asymmetries which are ratios of 
cross sections, are much 
less affected than simple cross sections by higher order corrections.
This behaviour has been confirmed recently
by some calculations who have provided some precise 
results on the small influence of gluons
on the standard
QCD-Electroweak interference term at NLO \cite{EMR}. 
A first estimate of the size of NLO corrections to $q-q$ scattering  
is in favor of a relatively small correction, of the order
of 10\% of the asymmetry itself \cite{BNL2001}.\\
Concerning experimental uncertainties, a good knowledge of 
the beam polarization ($\pm 5\%$) and a very good relative luminosity 
measurement ($10^{-4}$), should allow to get
a systematic uncertainty for
a single spin measurement of the order of 5\% \cite{BSSW}.
For the time being, we have taken into account all the present uncertainties
by using a global systematic error on the spin asymmetry 
$(\Delta A)_{syst}/A = 10\%$.\\

In Figures 1 and 2 we compare the non-standard asymmetries \alpp and \alnn
to the Standard one in each case, focusing on models $C$ and $D$.
We ignore here models 
$A$ and $B$ which don't give any effect on \alnn since 
in these models parity is conserved in 
the interactions between the corresponding
\z  and $d$ quarks (axial for Model $A$ and 
vector-like for Model $B$). The result of our calculation for these models
in the case of \ppv collisions has been already displayed in
ref. \cite{PTJMVZ'}.

\begin{figure}[htbp]
\vspace*{-2.1cm}
\begin{tabular}[t]{l r}
\epsfxsize=8cm
\epsfbox{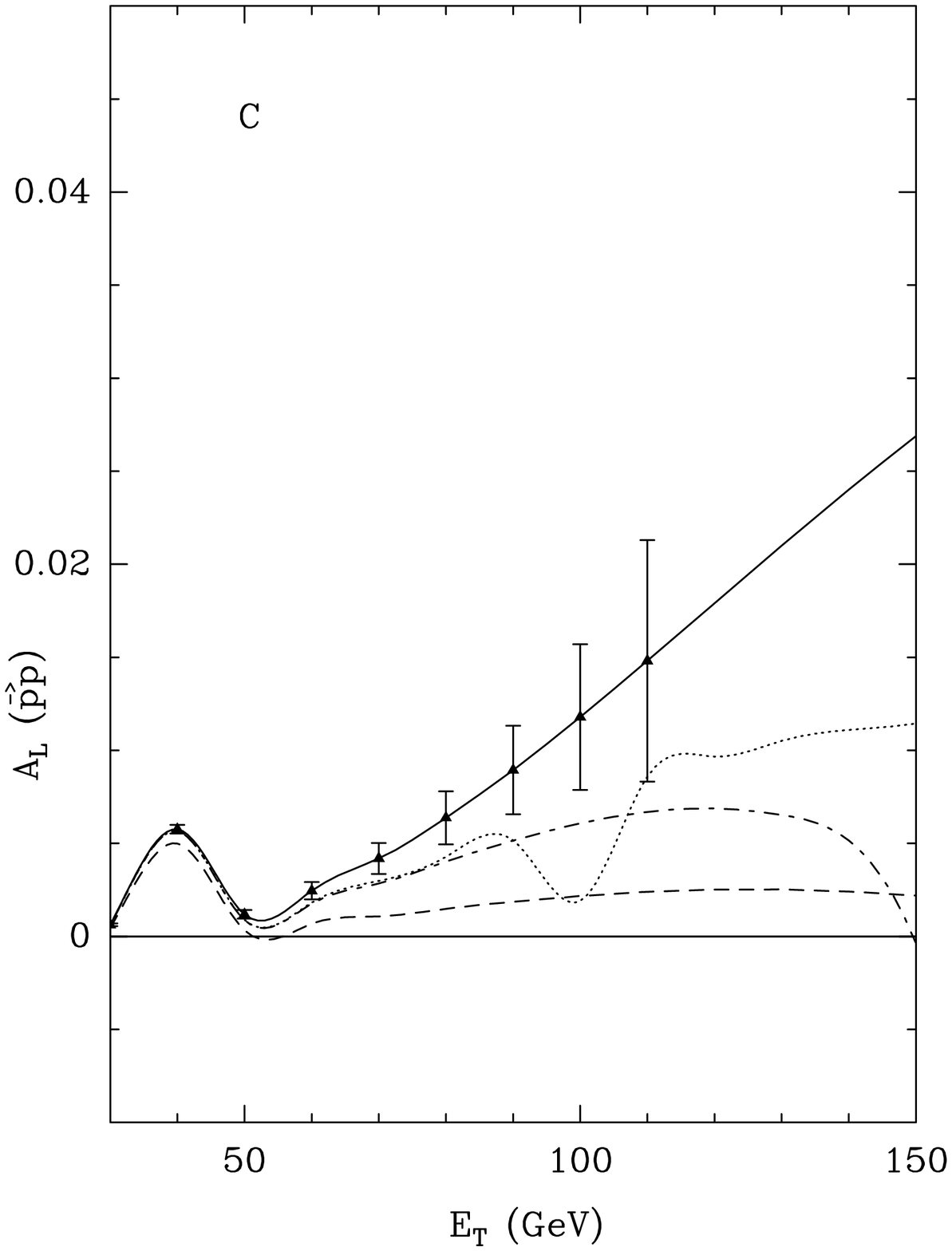}
\epsfxsize=8cm
\epsfbox{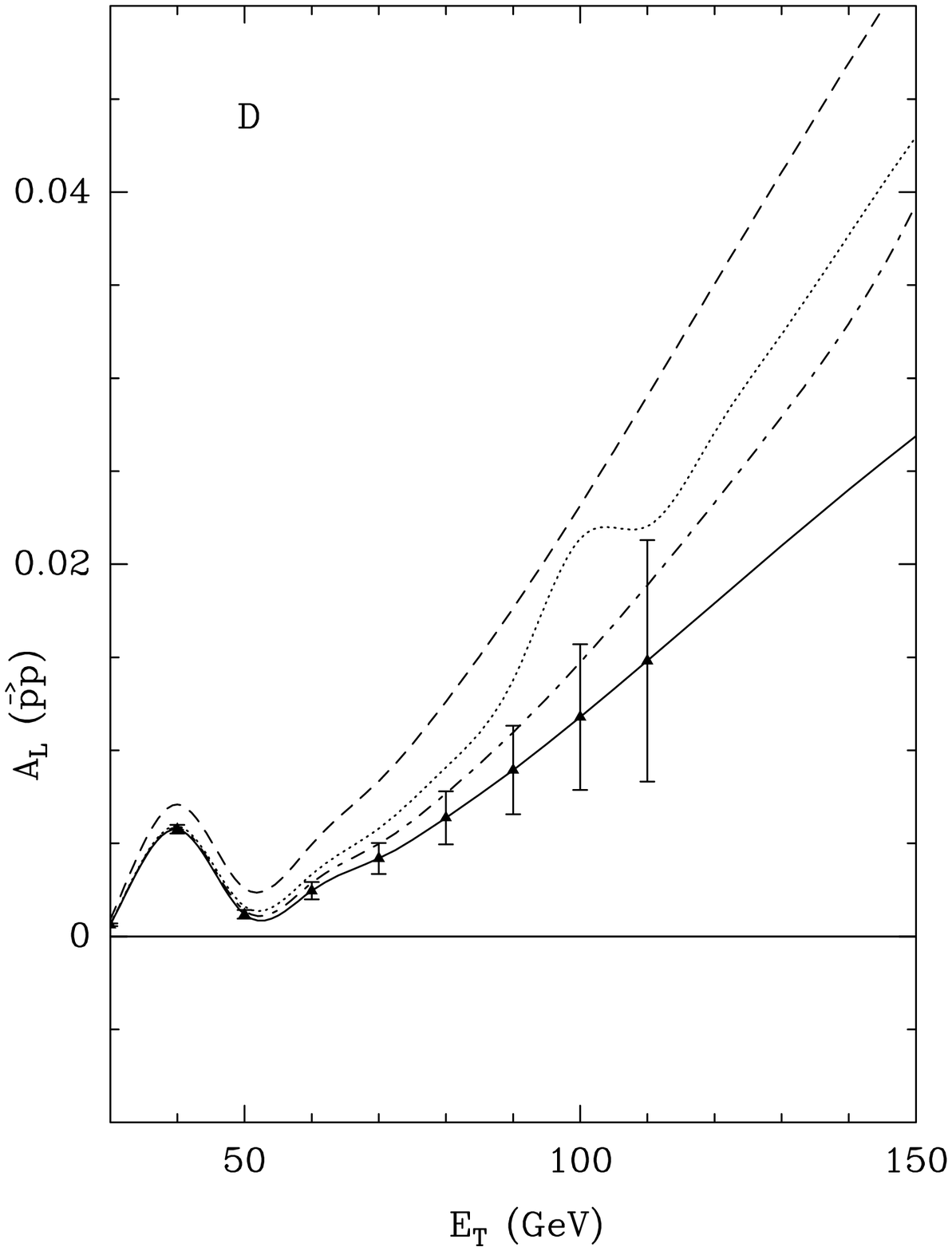}
\end{tabular}
\vspace{-1.4cm} 
\caption{$A_L$ versus $E_T$ for models $C$ (left) 
and $D$ 
(right) with $\vec{p}p$ collisions at RHIC
at ${\sqrt s = 500}$ GeV. The plain curves 
represent the SM predictions. The dashed, dotted and dash-dotted 
curves correspond to the cases where the masses are   
$M_{Z^{'}}=90,\;200$ et $300\;GeV$ respectively. The error bars 
correspond to the integrated luminosity 
$L_2=3.2\; fb^{-1}$. $\kappa =1$ for all cases except for the model
$C$ with  $M_{Z^{'}}=300\;GeV$ and $\kappa = 1.5$.}
\end{figure}
\begin{figure}[htbp]
\vspace*{-2.1cm}
\begin{tabular}[t]{l r}
\epsfxsize=8cm
\epsfbox{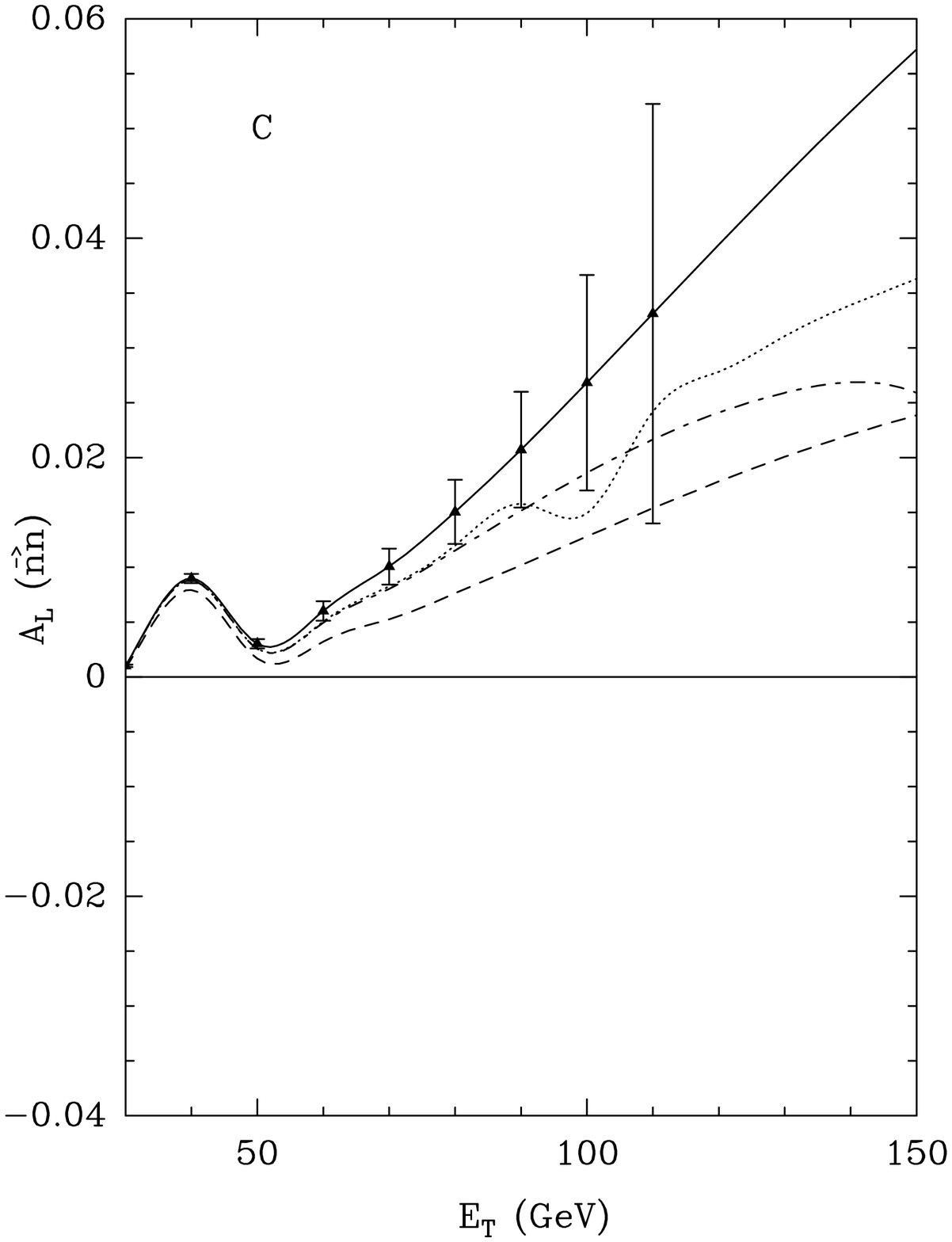}
\epsfxsize=8cm
\epsfbox{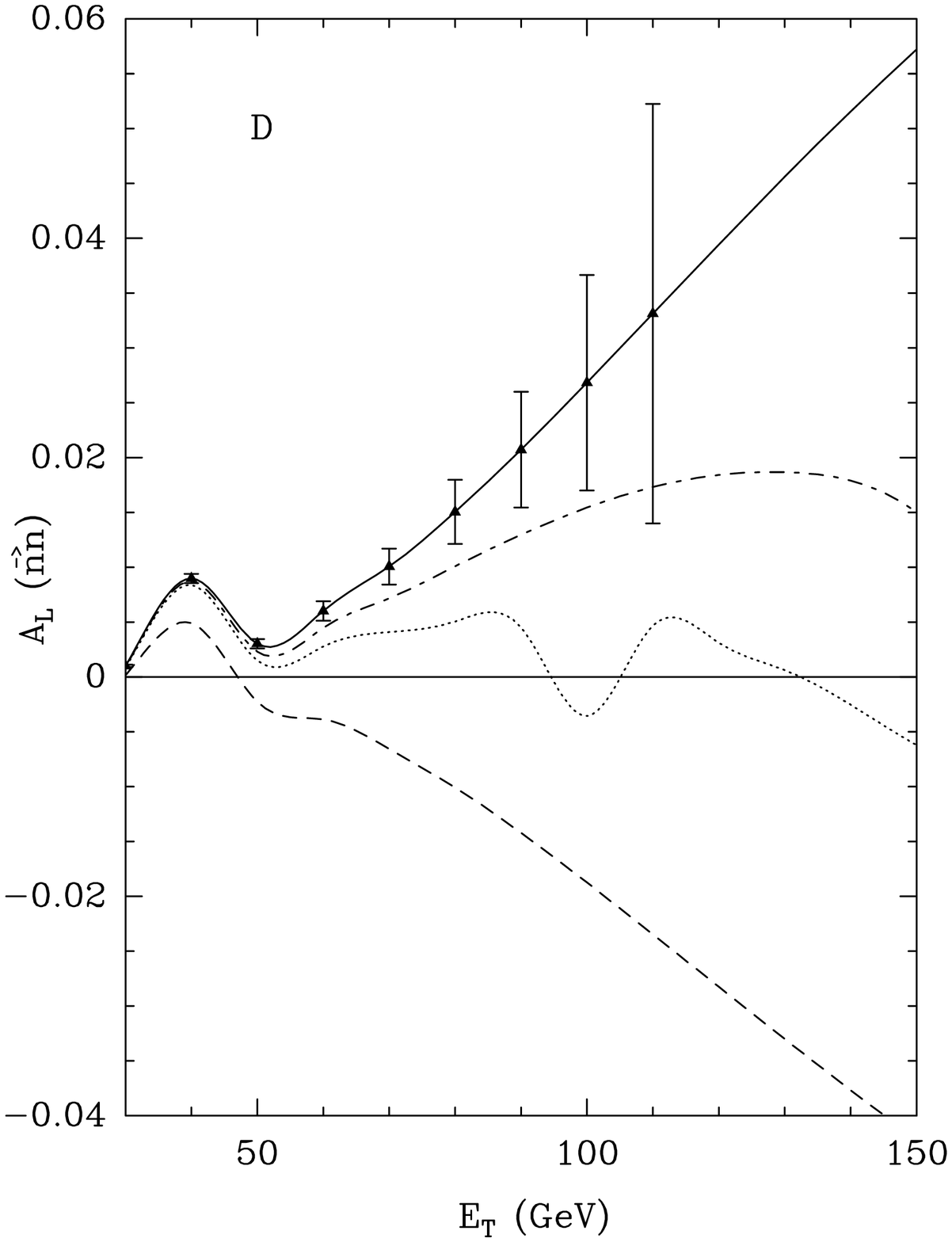}
\end{tabular}
\vspace{-1.4cm} 
\caption{Same as Fig.1 for $\vec{n}n$ collisions at  ${\sqrt s = 400}$ GeV.}
\end{figure}


Concerning the standard asymmetry, one can notice that, 
in spite of the smaller center of
mass energy the \alnn is larger than the
one  calculated for \ppv collisions in ref. \cite{BSSW,PTJMVup}
This is due to 
the larger parity violation in the $d$ quark sector compared to the $u$ quark sector
for  the SM ({\it i.e.} $|C_L^2-{C^d_R}^2|>>|C_L^2-{C^u_R}^2|$). However, 
the influence of the reduced energy shows itself on the error bars which are 
larger than for the ones on \alpp for the same $E_T$ bin.
The bumps in the curves correspond to the jacobian peaks
due to real $W$ and $Z$ exchanges around $E_T\approx M_{W,Z}/2$,
or $Z'$ exchange at $E_T\approx M_{Z'}/2$. The remaining effects 
on the whole $E_T$ spectrum are due to $Z.g$ and $Z'.g$ interference terms.

One can see that, in the framework of
the Models $C$ and $D$ the effects of the \z are
spectacular provided its mass is not too high, hence RHIC should not miss 
them if they are present. 

The deviations from the SM expectations are negative in Model $C$ in Fig.1
and Fig.2 in accordance with the right-handed dominance of both
$u$ and $d$ quark couplings.
On the other hand, in the case of Model $D$ which is dominantly
left-handed for $u$ quarks and right-handed for $d$ quarks, the deviation is
positive in $\vec{p}p$ collisions and negative in $\vec{n}n$ collisions.
\\
Finally, in the case of $\vec{n}n$ collisions,
if we compare the effects on $A_L$ coming from Models $C$ and $D$ for
the same $Z'$ mass, we see that Model $D$ implies some larger
deviations from the SM predictions. This difference is due to the larger
parity violation in the $d$ quark sector for Model $D$ compared to Model $C$.
Indeed, the difference $C_L^2-{C_R^d}^2$ appearing in eq.(\ref{ALs}) is equal
to 1/9 for Model $C$ and to $1/3$ for Model $D$, implying 
roughly a three times larger
deviation for model $D$.\\

In Figure 3, we present the limits
on the parameter space
$(\kappa,M_{Z^{'}})$  that both \alnn and \alpp should
provide with the integrated
luminosities  $L_1=800\;pb^{-1}$ and $L_2=3200\;pb^{-1}$ for models $C$ and $D$. 
We also display the inferred constraints coming from
the dijet cross section studies by the $p\bar{p}$ collider
experiments UA2  \cite{UA2}, CDF \cite{CDFjets2} and D0 \cite{D0jets}.
In fact the published results were restricted to the so-called
$Z'$ "sequential standard model" (SSM) with $\kappa = 1$. 
We have 
easily extrapolated
these results to models $C$ and $D$ by changing the couplings appropriately
for a reasonable range of $\kappa$ values. 
\begin{figure}[htbp]
\vspace*{-2.1cm}
\begin{tabular}[t]{l r}
\epsfxsize=8cm
\epsfbox{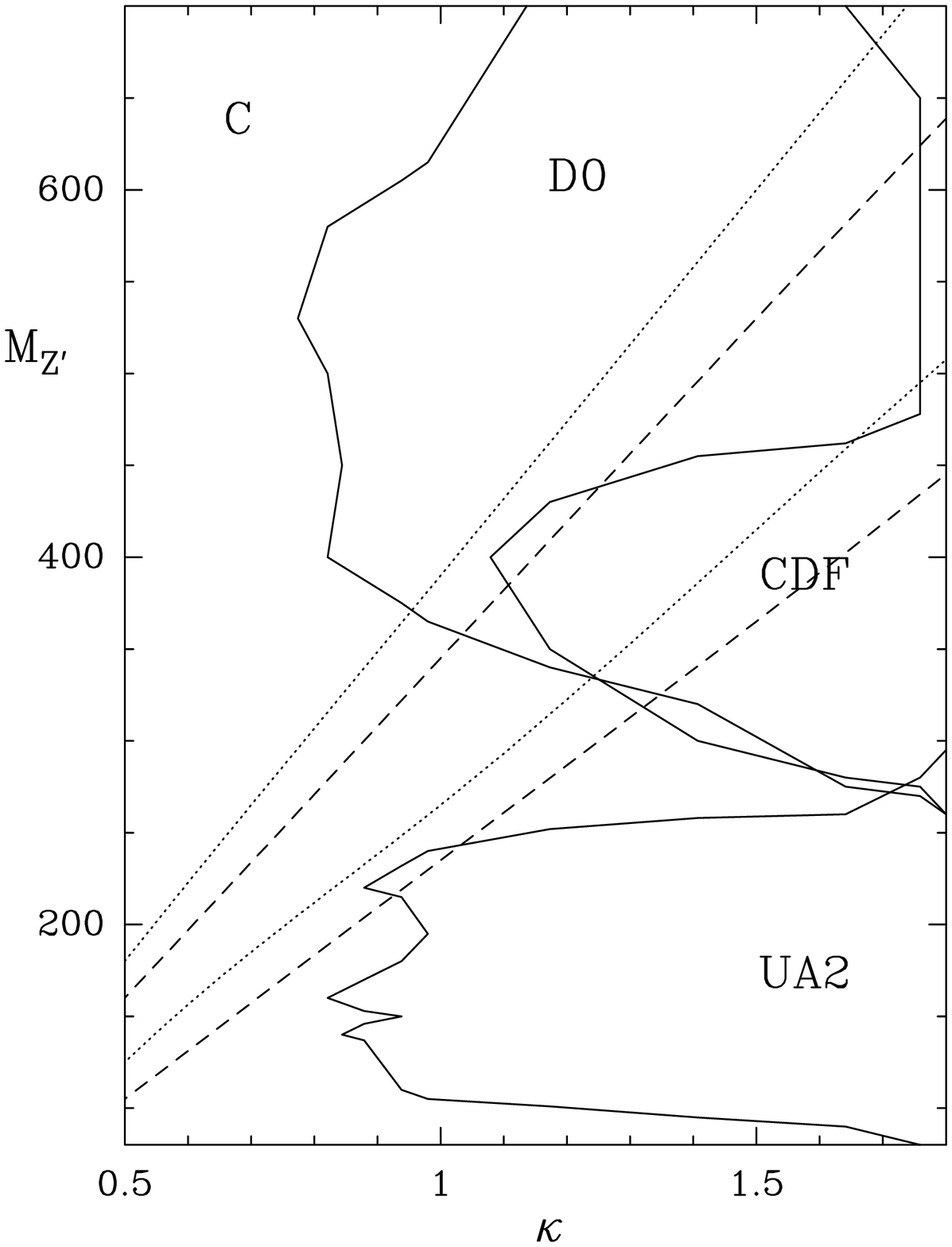}
\epsfxsize=8cm
\epsfbox{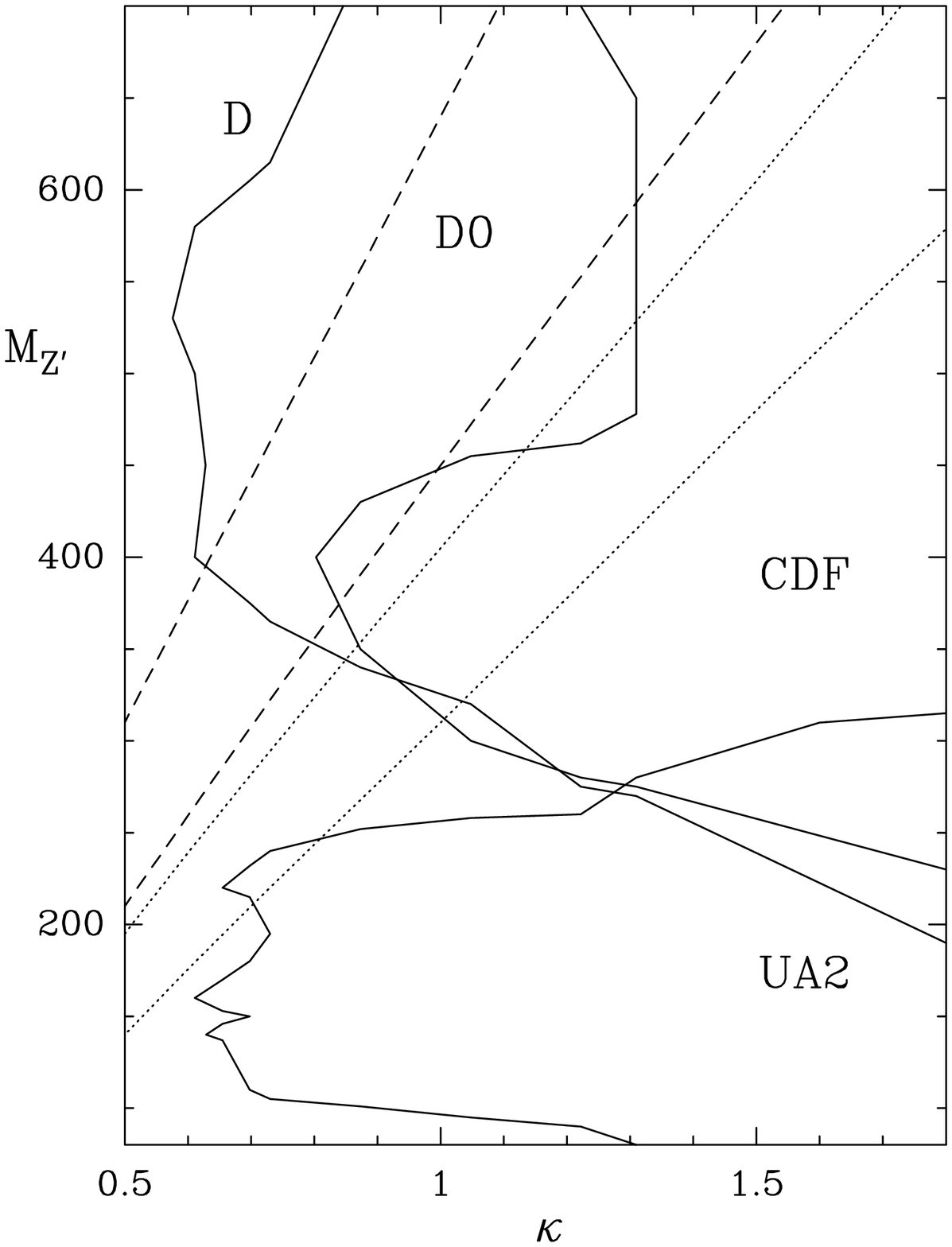}
\end{tabular}
\vspace{-1.4cm} 
\caption{Bounds on the parameter space $(\kappa,M_{Z^{'}})$ for models $C$ and 
$D$. The dotted and dashed curves correspond respectively to the predicted 
limits at 95\% C.L. 
from 
$A_{L}(\vec{p}\vec{p})$ et $A_{L}(\vec{n}\vec{n})$ at RHIC 
with $\sqrt{s}=500\, GeV$ 
for {\ppv}, $\sqrt{s}=400\, GeV$ for \nnv collisions, and with 
$L_1=800\;pb^{-1}$ (lower curves) 
and 
$L_2=3200\;pb^{-1}$ (upper curves).}
\end{figure}
 One can see that these unpolarized collider studies 
are not constraining a \ZP mass as soon as $\kappa$ is small enough.
Also, and this is true in any leptophobic model, 
some windows were still present around $M_{Z'}$ = 300 GeV$/c^2$ and below 
$M_{Z'}$ = 100 GeV$/c^2$. 
With the help of polarized hadronic beams the
situation could be greatly improved :
in particular the hole
centered on $M_{Z'} \approx 300 $GeV/$c^2$ should be covered 
provided $\kappa$ is greater than $\approx 0.7$.
One gets the same behaviour for the bounds in the framework of Models
$A$ and $B$ (see ref. \cite{PTJMVZ'}).
\\

We remark that \alpp is slightly more sensitive than \alnn for model $C$. This 
difference can be explained simply by the reduced center of mass energy for 
\nnv collisions. Indeed for the model $C$, the parity violation is equal 
in strength for both the
$u$ and $d$ quark sectors, {\it i.e.} 
$C_L^2-{C^u_R}^2=C_L^2-{C^d_R}^2$. 
Conversely, for Model $D$, we see that the sensitivity is clearly in favour of \alnn 
which can be understood thanks to the relation 
$|C_L^2-{C^d_R}^2|=3|C_L^2-{C^u_R}^2|$.\\

To conclude, the analysis of PV spin asymmetries measured within $\vec{n}n$
collisions is able to constrain the presence of a new weak hadronic
interaction in the $d$ quark sector. In the case of a discovery,
the deviations from the SM expectations indicate the chirality
of the new interaction in respect of $d$ quarks.\\

\section{Constraints on the scalar structure}

In this part, we want to analyze which kind of information could
be provided by the precise measurements of both spin asymmetries
at RHIC, namely \alpp and \alnn.
In particular, are we able to discriminate between the three scalar structures
we have described in Section 2 ?

At this stage it is worth recalling our assumptions : \\
{\it i)} the  condition of leptophobia plus a small $Z-Z'$
mixing angle, \\
{\it ii)} the gauge invariance under $U(1)^{'}$ of the fermion mass terms,\\
{\it iii)} $SU(2)_L$ invariance,\\
{\it iv)} we assume that some PV effects due to a leptophobic \z have been 
detected through the measurement of \alpp in the "first" phase 
of the RHIC operations
with polarized proton beam(s) : This means that parity is violated
in the $u$ quark sector of the new  $U(1)^{'}$.\\

In figure 4, we present \alpp versus \alnn for a transverse energy 
$E_T=70\pm  5\;GeV$. We have chosen this particular interval since its 
contribution to the $\chi^2$ function involved in the analysis is maximal.
 Of course,
a full integration over the $E_T$ range accessible experimentally will
reduce the error bars. However, to be realistic, this integration should
take into account the details of the jet reconstruction of the RHIC detectors,
an analysis which is far beyond the scope of this paper.\\
\begin{figure}[htbp]
\begin{center}
\vspace*{-3cm}
\begin{tabular}[t]{l r}
\epsfxsize=12cm
\epsfbox{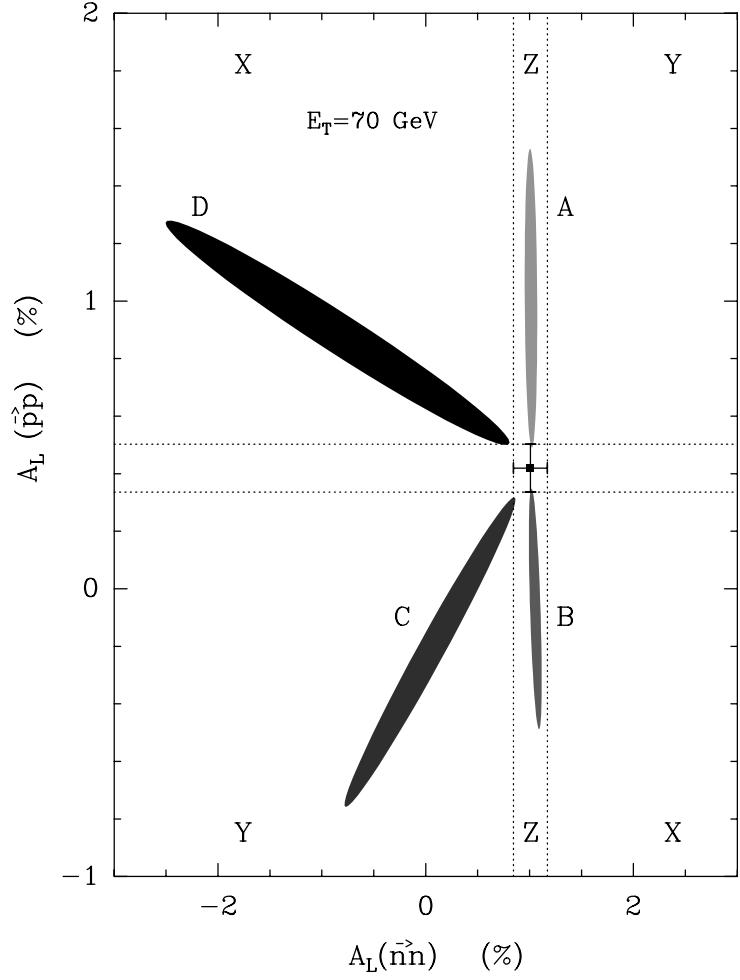}
\end{tabular}
\end{center}
\vspace{-1cm} 
\caption{Predictions of the various models (see text)
in the plane $A_{L}(\vec{p}p)$ versus 
$A_{L}(\vec{n}n)$ for 
$E_T=70\;GeV$. The error bars on the SM point
correspond to the integrated luminosity 
$L_2=3.2\; fb^{-1}$.}
\end{figure}
In this figure, the central point represents the SM 
prediction. The error bars correspond to the integrated luminosity 
$L_2=3200\;pb^{-1}$.
The models  $A$, $B$, $C$ and $D$, which were introduced to 
illustrate the different
scalar structures, are represented each one by a shaded ellipse.
Each point inside an ellipse corresponds to a precise value
of $\kappa$ and $M_{Z^{'}}$, these values are taken as to satisfy the present
experimental constraints presented in Fig.3 and in Fig.1 of Ref.\cite{PTJMVZ'}.\\
Concerning the shapes of the ellipses, 
there is no simple dependence on
the two parameters $\kappa$ and $M_{Z^{'}}$ in this plane. 
However,  a point close to the "SM cross" obviously means that
one has a small $\kappa$ and/or a large $M_{Z'}$.
Conversely, the biggest effect (that is the farthest from the SM point)
is obtained for $M_{Z'} = M_Z$ (which is our lowest $M_{Z'}$
value) and for the highest experimentally allowed value of $\kappa$ within each
model.\\

Remember that {\it structure I} is characterized by a vector-like
coupling of the $d$ quark to the \ZP. Models 
belonging to {\it Structure II}
should exhibit the same left-handed or right-handed dominance for
$u$ and $d$ quarks. It means that the deviations 
from SM expectations for \alpp and for \alnn 
go in the same direction.
In the framework of {\it structure III}, corresponding to highly
non-trivial scalar structures, no predictions are made 
for the $d$ quark couplings, hence
they can be located anywhere in the plane 
(\alpp,\alnn). 
However, it is only for this structure that we can have an opposite chirality for
$u$ and $d$ quarks couplings. 
On fig.4 we call "X" the two regions which correspond to the latter case,
in the upper-left and lower-right sectors. Model $D$ 
is an illustration of this situation.
A first conclusion is that experimental results belonging to zone "X"
should allow to eliminate both {\it structure I} (i.e. the 2HDM) and 
{\it structure II}
whose common property is the presence of trilinear mass terms 
for $u$ and $d$ quarks .\\

Secondly, we define sector "Y" which is accessible by models
from {\it structure II} or {\it structure III} but wich excludes
{\it structure I}. Concerning {\it structure II}, the fact that the points
belong to zone "Y"
is related to the common property of left-handed dominance (right-handed dominance)
for $u$ and $d$ couplings, corresponding to the
upper-right (lower-left) sector  of the plane. Model $C$ belongs to this
category with a right-handed dominance.

Finally, 
for the models of {\it structure I}, like model $B$, 
the $d$ quarks have some vector-like couplings 
to the \z and we don't expect any deviation on
\alnn. So, they should be located on the vertical line passing through the 
SM point. Taking into account the experimental conditions, this 
line is replaced by the band "Z" whose width is determined by the error on the
standard \alnn.  
Of course models belonging to the two other structure can 
fall into this band (Model $A$ is an illustration from {\it structure III}).
Conversely models from {\it structure I}, which characterize the 2HDM with
trilinear mass terms for all fermions,
will be ruled out by the RHIC \pp and \nn 
collision experiments if any effect is observed outside this band.\\
On the other hand, if for some other phenomenological or theoretical
reason, it turns out that models from {\it structure III} have to be rejected, then
the "Z" band should be a clear signature of the simplest 2HDM's 
of {\it structure I}. The only drawback is the case of axial couplings
of the $d$ quarks which is forbidden by these models but is allowed as
a very particular case of the models of {\it structure II} and which
contaminates the "Z" band.

\section{Conclusion}

The existence of a new weak force between quarks 
at a not too high energy scale is an attractive
possibility which is not ruled out by present data. 
If a new  neutral gauge boson \z owns the property
of leptophobia, it evades the bounds
coming from LEP/SLC experiments and it must be looked for in
purely hadronic processes.

The polarized proton beams which are available at RHIC allow the
precise measurement of the PV spin asymmetry $A_L$ 
in the production of jets. 
As pointed out in our previous papers, and also stressed by some authors
\cite{LopezNanopoulos}, such measurements could really 
lead to a discovery if the new \z exhibits some handed couplings to $u$ quarks,
since $u$ quarks play a dominant role in the collision process. 
In addition, as usual, measuring a spin asymmetry allows to get 
a handle on the chiral structure of the underlying interaction.
More precisely, the sign of the
deviation from the expected standard value of
$A_L$ should allow to pin down the chiral structure of the 
new interaction, still in the $u$ quark sector.\\
In spite of its relatively low center of mass energy
(500 - 600 GeV) the RHIC machine should be a remarkable tool, in particular
thanks to the very high luminosity which is expected. Hence, in some models a mass
as high as $M_{Z'} = 400$ GeV could give a measurable effect.

Since the acceleration and storage of high intensity polarized
$^3He$ ions, which means polarized neutrons, will be a real possibility
in a second phase of RHIC, it is valuable to investigate what could be obtained
on the $d$ quark sector.

We had already checked that polarized {\it proton-neutron} collisions could only give
access to the effects of a new charged current \cite{PTJMVW'}.
For testing the $d$ quark sector of a new $U(1)'$, the use of both "neutron
beams" is mandatory.

In this paper, we have checked first
that it would be possible to get some valuable information on  
the chirality of the $Z'd\bar{d}$ vertex
thanks to the measurement of the asymmetry \alnn.
This could be done with a precision which is comparable to what can
be hoped from the measurement of \alpp.

Moreover, getting in the same time an information on the $u$ and the $d$ couplings
is a way of testing the scalar structure of the underlying model.
We have seen  that the property of leptophobia, plus some general assumptions
like gauge invariance under the standard $SU(2)_L$ and 
under the new $U(1)'$, constrains the Higgs sector of the model.

The simplest case of 2-Higgs Doublet Models, with trilinear mass terms and
the traditional property $H_d \equiv H_l$, exhibits the interesting consequence
of vector-like couplings of the \z to $d$ quarks along with PV violation in the
$u$ quark sector in general. 
Other models exhibit a
more elaborated scalar structure, in particular the scalar sector which gives masses
to ordinary quarks could be decoupled from the corresponding sector for
leptons. This is the case in our models from {\it structure II} where we have 
considered two Higgs doublets giving masses to $u$ and $d$ quarks along 
with some phenomenological constraints, without any assumptions on the leptons. 
In this
case parity is violated in general in both the $u$ and $d$ sectors.
Measuring \alpp and \alnn should allow to separate the two structures, if
an even more general possibility was not present. 
Conversely, if one does not assume anymore
the presence of trilinear mass terms for the quarks ({\it structure III}), the situation is
more open in the plane (\alpp,\alnn). Therefore, some definite conclusions
could be yielded only if the measurements are in favor of 
the "sector X" described in section 4 : In this case
{\it structures I and II} are forbidden. 
Similarly an effect observed outside the "Z band" rules out
 the 2HDM's ({\it structure I}) without any ambiguity.

Our conclusion is that the implementation of polarized "neutron beams" 
at RHIC should greatly complement the program of New Physics searches
with polarized proton beams, since a non trivial piece of information
could be obtained on the scalar sector of the underlying theory.

\vspace{1.cm}

\no {\bf Acknowledgments}\\ 
E.T acknowledges Prof. Y. Yayla, previous president of the Galatasaray University, 
to have supported his studies at the Centre de Physique Th\'eorique, CNRS-Marseille.\\
J.M.V. acknowledges the warm hospitality at the RIKEN-BNL Research Center where
part of this work has been performed. Thanks are due to 
G. Bunce, G. Eppley, N. Saito, J. Soffer, M. Tannenbaum and W. Vogelsang 
for fruitful discussions.



\end{document}